# Production of CO$^{3+}$ through the Strong-Field Ionization and Coulomb explosion of Formic Acid Dimer

Shaun F. Sutton[1,2], Dane M. Miller[1,2], Lenin M. Quiroz[1,2], Ananya Sen[1,2], Pilarisetty Tarakeshwar[1], and Scott G. Sayres[1,2]*

1 School of Molecular Sciences, Arizona State University, Tempe, AZ 85287
2 Biodesign Center for Applied Structural Discovery, Arizona State University, Tempe, AZ 85287

Femtosecond laser pulses are utilized to drive multiple ionization in gas-phase formic acid dimers and ions are studied using time-of-flight mass spectrometry. The interaction of formic acid dimer with 200 fs linearly polarized laser pulses of 400 nm with intensities up to 3.7x10$^{15}$ W/cm$^2$ produces a carbon monoxide trication. Experimental measurements of the kinetic energy release of the ions are consistent with our molecular dynamics simulations of the Coulomb explosion of a formic acid dimer and suggest that no movement occurs during ionization. KER values were recorded as high as 44 eV for CO$^{3+}$. Potential energy curves for CO$^{n+}$, for n ≤ 3, have been calculated using CCSD(T) and confirm the existence of metastable states with a large potential barrier with respect to dissociation. This combined experimental and theoretical effort shows the existence of sufficiently long-lived CO$^{3+}$.

Strong-field laser pulses have become an essential tool used at the forefront of experimental atomic, molecular, and optical physics. The process of strong-field ionization (SFI) lies at the heart of modern ultrafast spectroscopy. Not only is SFI the underlying mechanism responsible for high-harmonic generation (HHG) [1,2] used in extreme ultraviolet (XUV) [3–5]; it has also become the dominant technique for the production of coherent electronic wave packets that operate on the attosecond timescale [6,7]. Initial ionization of atoms and molecules are well described under the single active electron (SAE) approximation, where only the outermost electron is affected by the external laser field. Non-sequential double ionization (NSDI) is a phenomenon that has received a lot of attention where the ionization rate cannot be interpreted by a single electron and is typically driven by recollision of the freed electron. NSDI dominates below the saturation intensity [8]. The ionization behavior of molecules is influenced by dynamic changes that are initiated by the laser field leading to several enhanced ionization mechanisms and phenomenon. In diatomic molecules, the ionization rate can be strongly enhanced around a critical internuclear distance (CREI) [9]. This enhanced ionization has been suggested for triatomic [10] and polyatomic molecules. However, how bonding affects the ionization rates in polyatomic molecules, especially beyond the first electron, is not resolved. Beyond the valence electron shell, ionization proceeds back to atomic-like in molecules [11,12]. Thus, polyatomic molecules have an important role to play in advancing strong-field physics by revealing details about electron correlation as it is modulated by nuclear motion and bonding.

Fundamental questions regarding the structures, stabilities, and bonding schemes of small multiply charged ions test our concepts of chemical bonding. Although larger molecules can stabilize multiple positive charges by spreading them across large distances, multiply charged diatomic molecules are exotic gas-phase species and require high amounts of energy for production. Intense fs lasers, X-ray lasers, ion collisions, and electron impact all provide access to multiply charged molecular cations. Typically, repulsion between the charges drives dissociation or Coulomb explosion (CE) of the ions, but if a sufficiently deep potential well exists, a multication can be produced even if the energy of the bound state exceeds the energy of the individual ions (metastable well). Such multiply charged ions play important roles in high energy environments, such as planetary atmospheres [13–16] and interstellar clouds [17,18].

CO is the second most abundant molecule in interstellar space making its high charge states particularly important [19–25]. CO$^{2+}$ was first identified in 1931 [26] and has since been analyzed in many experimental [27–29] and theoretical studies [30–36]. The cross-sections and multiple ionizations of CO$^{n+}$ (as high as n=9) have been studied with a variety of techniques, including electron impact, [37,38] fast ion collisions [39–42], VUV photoionization [34,43,44], MPI [45,46], and strong-field ionization [47,48,57,49–56]. The lifetime of metastable CO$^{2+}$ ranges from a few tenths of a nanosecond to several seconds [58,59]. Yet, the direct observation of CO$^{3+}$ has so far eluded mass spectrometry experiments, suggesting that ions with n > 2 dissociate within 300 ns [41]. Trications require higher excitation energies for their formation and must resist higher Coulomb repulsion forces, making them less stable than dications but also more interesting.

In this Letter, we study the formation of CO multications from the ultrafast strong-field ionization of larger polyatomic molecules. We demonstrate that the trication of CO contains a metastable state and can be accessed directly with femtosecond laser pulses. In this combined experimental and theoretical effort, we employ fs laser pulses to strong field ionize formic acid dimers and use mass spectrometry to record multications. The Coulomb explosion (CE) of formic acid dimer produces CO$^{3+}$.

Our calculations reveal the potential energy curves of $CO^{3+}$ are metastable. Molecular Dynamics (MD) simulations are compared to the measured kinetic energy release (KER) from the CE.

Formic acid (HCOOH), seeded in He, is pulsed into a high vacuum chamber, where the gas traverses a skimmer to produce a molecular beam and travels 10 cm before reaching the extraction region of a Wiley-McLaren [60] time-of-flight mass spectrometer. Formic acid is well known to dimerize, with a binding energy of ~0.6 eV [61]. Ionization of the neutral molecules occurs through interaction with a synchronized fs laser pulse. The ions are accelerated toward the detector using electrostatic grids (Fig. 1). The Ti:Sapphire laser system outputs 10 mJ per pulse at 800 nm, and operates at 30 Hz. The pulse width of the 800 nm laser beam was determined through an autocorrelation to be <200 fs. The 800 nm beam passes through a BBO crystal for second harmonic generation and 2 mJ of linearly polarized 400 nm light is directed into the chamber using a 50 cm focal lens, to reach a laser intensity of $3.7 \times 10^{15}$ W/cm$^2$ and drive tunneling ionization and fragmentation of the molecular beam. Additional details are provided in the Supplemental Information.

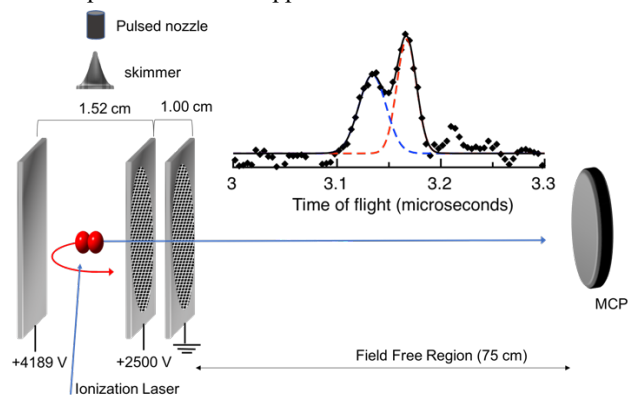

Fig. 1. Mass spectrometer and Coulomb explosion measurements. The blue ion fragment is sent toward the detector, whereas the red fragment is sent backwards and is turned around thereby arriving at a later time. Time-of-flight peak splitting for $CO^{3+}$ is presented in the inset, where the 44 eV KER produces two peaks split at a flight time of ~3.15 μs that corresponds to m/z = 9.33 amu.

Fragment ions of $(HCOOH)H^+$, $COOH^+$, and $H_3O^+$ are produced via the proton-transfer reaction and dissociative ionization of $(HCOOH)_2$. The major fragments recorded in the mass spectra are $HCO^+$, $HCOOH^+$, and $COOH^+$ with minor products $CO^+$ and $CO_2^+$ (Fig. 2). Signals corresponding to $COO^+$, $HCOO^+$, and $HCOOH^+$ are broadened in arrival time, indicating KER caused by CE. The dissociation pattern is in agreement with previous reports of formic acid in intense laser fields [62,63]. In addition to heavy fragmentation of the molecule, the high laser intensities used in this experiment, $3.7 \times 10^{15}$ W/cm$^2$, produces multiply charged C, O, and CO ions (Fig. 2). The focus of this Letter is the production of multiply charged CO ions appearing at m/z = 28, 14, 9.33 that arise from the multiple ionization and CE of formic acid dimer.

Detailed *ab initio* calculations using the aug-cc-pVQZ basis sets were carried out at the CCSD(T) {coupled-cluster with single, double, and perturbative triple substitutions} levels of theory on multiply charged $CO^{z+}$ (z = 0-3) cations [64,65]. The electron configuration of CO contains an outermost orbital $(5\sigma)^2$ as virtually a nonbonding orbital, while the next two orbitals $(4\sigma)^2$ and $(1\pi)^2$ are strongly bonding orbitals. The ground state of CO is $^1\Sigma^+$, having an equilibrium bond length, $r_e$, of 1.128 Å. Our calculated potential energy curves (SI Fig. S1 and Table S1) are in excellent agreement with prior results. In particular, our calculations predict the ground electronic states of $CO^{2+}$ are metastable, where the $^3\Pi$ and $^1\Sigma^+$ states are nearly degenerate, with the $^3\Pi$ being slightly lower in energy, in agreement with previous reports. The binding energy is 4.7 eV for the ($^1\Sigma^+$) dication and has a bond length of 1.156 Å, in agreement with previous calculations of ~5 eV [66]. The dissociation energy of the $^3\Pi$ is 1.77 eV, and has an extended equilibrium bond length of 1.249 Å [35,58,66,67]. The vertical IP for the $^3\Pi$ electronic state is calculated to be 41.7 eV in close agreement with the measured appearance potential of $CO^{2+}$ of 41.294 ± 0.010 eV [25]. The excited $^3\Pi$ states are well separated meaning shallow well is a consequence of a delicate balance between the Coulombic repulsion and covalent interaction between a $C^+$ and $O^+$. However, the excited $^1\Sigma^+$ states are nearby, producing avoided crossings that lower the binding energy of the state.

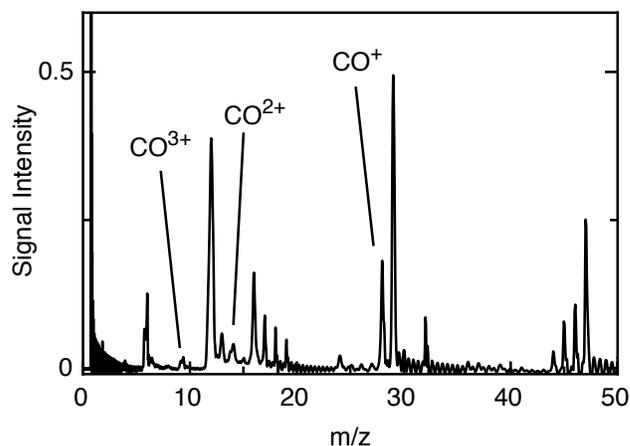

Fig. 2. Mass spectrum (1000 averages) of the CE of formic acid dimer normalized to the most intense peak (H$^+$). The peaks attributed to multiply charged CO are labeled. The intensity of the dimer (m/z = 92) is 0.03 on this scale (not shown).

Previous calculations suggest that $CO^{3+}$ is purely dissociative [33,36,68]. In agreement with literature results, we find that both the $^2\Pi$ and $^4\Sigma^+$ ground states have no local minimum along the CCSD(T)/aug-cc-pVQZ potential energy curves and are therefore purely dissociative. The $^4\Sigma^+$ This is due to a weak interaction between closed shell species $C^{2+}$ ($1s^22s^2$ $^1S$) and $O^+$ ($1s^22s^22p^3$ $^4S$). The potential energy curves for the trication are shown in Fig. 3. The $^2\Pi$ symmetry (corresponding to $C^{2+}$ ($^1S$) + $O^+$ ($^2D$)) for $CO^{3+}$ contains several closely lying excited states with the same symmetry, resulting in several avoided crossings, [36] where breakdown of the Born Oppenheimer approximation may allow non-adiabatic electronic transitions and enable a metastable $CO^{3+}$ configuration. At Franck-Condon overlap of CO ($r_e$ = 1.128), we calculate a vertical IP of ~82.0 eV, in agreement with the threshold for triple photoionization of CO measured using vacuum ultraviolet (VUV) radiation of 81 ± 2 eV [34,43,44]. However, we also find that both states contain shallow metastable excited states at ~83.6 eV. Similar to the dication, the higher spin state for the trication is slightly lower in energy and has a longer equilibrium bond length (Fig. 3). A shallow $^2\Pi$ well is located with an equilibrium bond length of

1.122 Å, and a well depth of 4.50 eV. Near the minimum of the $^2\Pi$ state an avoided crossing exists with the purely dissociative Coulomb potential. An excited $^4\Sigma^+$ state has an equilibrium bond length of 1.296 Å and well depth of 0.49 eV. The bound $^4\Sigma^+$ state is isolated from the curve crossing that occurs in the $^2\Pi$ state.

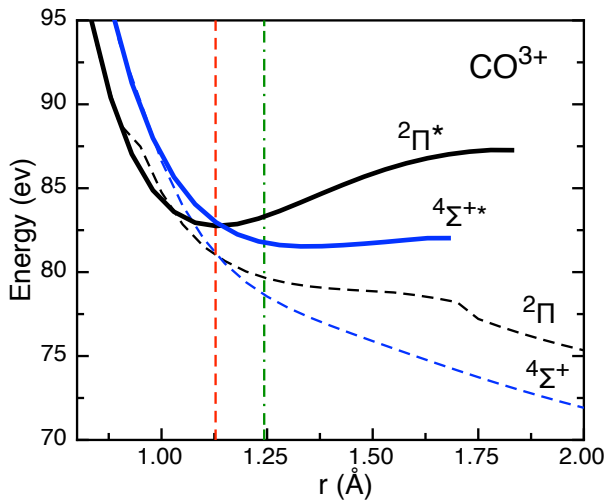

Fig. 3. Calculated potential energy curves for $CO^{3+}$. These curves are given in energy relative to the minimum of the $X^1\Sigma^+$ electronic ground state of neutral CO. The vertical lines represent the bond length of molecular CO and in formic acid.

Upon exposure of formic acid to intense lasers pulses, multiple ionization leads to severe fragmentation and CE. However, the prominence of $CO^{n+}$ in our spectrum indicates that it has a strong tendency to remain intact through the ionization process. This is distinctly different from when ionization of formic acid is driven with picosecond duration laser pulses, where $CO^{2+}$ was not observed even when atomic high charges states are observed [45,46]. Previous experiments indicate that the $CO^{2+}$ yield was observed to be similar to that of the $C^+ + O^+$ channel with femtosecond pulses [56]. Thus, femtosecond laser are required for the ionization process to outpace dissociation and complete atomization when intense laser beams are applied. Femtosecond lasers reach extreme laser intensities, where the electric field of the laser becomes comparable to the electron Coulomb binding potential. The ionization mechanism transitions from multiphoton ionization to tunneling ionization for values << 1 according to the Keldysh parameter, $\gamma=\sqrt{(IP/2U_P)}$, where IP is the ionization potential and $U_P$ is the laser's ponderomotive potential. The Keldysh parameter is 0.32 at the $3.75\times10^{15}$ W/cm$^2$ employed here. For formic acid (IP=11.33 eV [69]), electron tunneling drives the initial ionization at laser intensities > $3.8\times10^{14}$ W/cm$^2$. The laser intensity necessary for the production of each species was calculated assuming tunneling ionization. The well-known Ammosov-Delone-Krainov (ADK) model [70] is applied to calculate the ionization rate and probabilities (Fig. 4) for each atom using knowledge of its effective quantum numbers, IP and laser conditions.

In our experiment, the magnitude of the $C^+$ signal is double that of the $O^+$ signal. The difference in the signal intensity between these two ions is expected from the difference in their IPs and the exponential dependence of tunneling with IP. Fragmentation of cation states leaves the electron on the atom with the higher IP. A prominence of $C^{2+}$ (~20x) relative to $O^{2+}$ is recorded. The appearance intensity of $C^{2+}$ is ~$2\times10^{14}$, whereas the $O^{2+}$ is ~4x higher. Multiply charged CO ions are prominent in the ion distribution. In fact, the signal for $CO^{2+}$ and $CO^{3+}$ are both larger than $O^{2+}$ in the mass spectrum (Fig. 2). The atomic ADK ionization rates can also be applied to understand the ionization sequence and suggests that $C^{2+}$ and $CO^{2+}$ arise at similar intensity and lower than $O^{2+}$. Thus, either the C or the CO fragment achieve double ionization before the O. This explains the double ionization of CO observed with few-cycle laser fields, where it is determined that $O^+$ arises with similar laser intensity as $CO^{2+}$ [57]. The CO dication has been well established to have long lifetimes, lasting into the seconds timescale and greatly exceeding the laser pulsewidth. This enables sequential ionization to reach the trication state. In our experiment, intact ions are recorded only if they are bound for at least the time spent within the extraction region of the mass spectrometer, placing a minimum lifetime of the $CO^{3+}$ lifetime at 166 ns (see SI). The ultimate fate of the trication is to dissociate to $C^{2+}$ and $O^+$ due to their difference in IPs. The high IPs for the $C^{3+}$ and $O^{3+}$ limits their production under the laser intensities employed here. In contrast, $CO^{3+}$ requires a laser intensity of ~$8\times10^{14}$ W/cm$^2$.

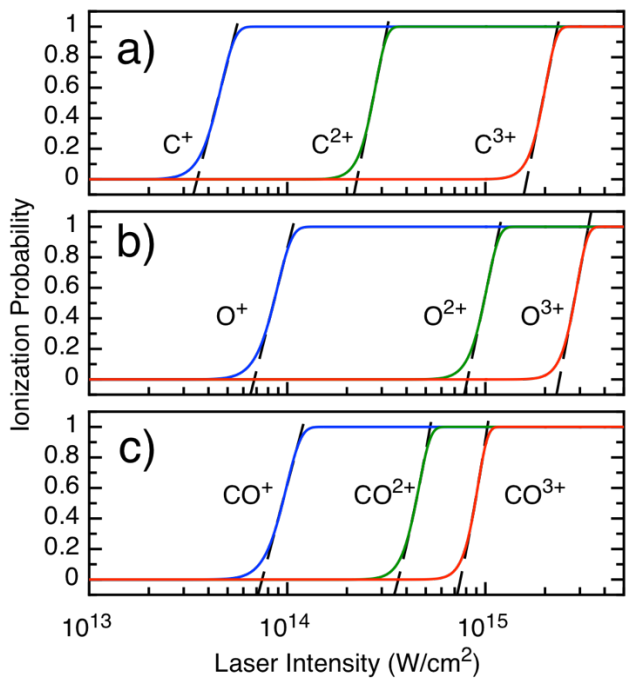

Fig. 4. ADK ionization probabilities for a) $C^{n+}$ b) $O^{n+}$ and c) $CO^{n+}$, with a 200 fs square wave laser pulsewidth.

Photoionization can bring about the formation of di- and trications, however the yields are often low due to unfavorable Franck-Condon factors in the vertical transition between the neutral molecules and the corresponding ion states. For example, KER values were measured that correspond to tetracation states from few-cycle experiments of CO, yet only the dication was observed [57]. Often, removal of electrons leads to an extended bond length in comparison to the neutral molecule. For example, the equilibrium bond length of molecular CO is too short and couples high in the trication $^4\Sigma^+$ potential well with sufficient energy to dissociate and has limited its observation. In contrast, the extended C=O bond length present in formic acid ($r_e$ = 1.243 Å), as opposed to in CO ($r_e$ = 1.128 Å), is critical because it enables better Franck-Condon overlap with the higher spin states of both the dication and the trication, which are lower in energy than the low spin states. Thus, upon sequential ionization of formic acid, there is no driving force for the bond to expand and the $CO^{2+}$ and $CO^{3+}$ are produced with a colder vibrational level and can remain bound.

Further insight to the $CO^{3+}$ production mechanism is obtained by considering the KER of the ions produced through CE. SFI of formic acid dimer occurs much faster than the nuclei can adequately respond, leading to the production of multiply charged ions in close proximity. The forces between the ions generates a significant KER and is measured with the mass spectrometer using the peak splitting method (Fig. S2) [71]. The measured KER values are presented in Table 1. Given the large number of avoided crossings among the electronic states of the multications and the important role played by the non-adiabatic couplings in the photodissociation dynamics, an assumption of a purely Coulombic repulsion for the KER is reasonable.

MD simulations for the formic acid dimer produce KER values in agreement with experimental measurements (Table 1). The KER results reveal that the molecule, and in particular the CO bond, did not undergo substantial movement or expansion during ionization. This suggests that the ionization rate is sufficient to produce highly charged ions before they have time to move and therefore samples the neutral's geometry. Due to the $C_{2h}$ symmetry of FAD, both C atoms obtain identical KER values from the MD simulation. However, there are two unique O atoms in each formic acid molecule. Simulations of the FAD predict that O ion from C=O obtains ~50% more KER (SI Table 1) than the O-H. For example, assuming all ions are +1, the $O^+$ obtains 12.8 and 17.9 eV. Experimentally, we cannot distinguish between the two different oxygen atoms in the mass spectra, and so they are averaged together.

**Table 1. Experimental and average MD KER values for the high charge states and the minimum laser intensities required for the ionization threshold.**

| Ion | Measured KER (eV) | MD[a] KER (eV) | MD[b] KER (eV) | IP (eV) | I (W/cm$^2$) |
|---|---|---|---|---|---|
| $C^+$ | 17.7 | 18.0 | - | 11.3 | $3.5\times10^{13}$ |
| $C^{2+}$ | 85.8 | 75.7 | - | 24.4 | $2.3\times10^{14}$ |
| $C^{3+}$ | - | 173 | - | 47.9 | $1.6\times10^{15}$ |
| $O^+$ | 13.0 | 15.3 | 12.6 | 13.6 | $6.8\times10^{13}$ |
| $O^{2+}$ | 47.8 | 58.9 | 49.0 | 35.1 | $8.0\times10^{14}$ |
| $O^{3+}$ | - | 130.4 | 109.2 | 54.9 | $2.4\times10^{15}$ |
| $CO^+$ | 1.4 | - | 4.8 | 14.0 | $7.5\times10^{13}$ |
| $CO^{2+}$ | 22.1 | - | 19.8 | 28.1 | $3.7\times10^{14}$ |
| $CO^{3+}$ | 44.4 | - | 45.326 | 39.9 | $7.5\times10^{14}$ |

a) The MD simulation is performed assuming atomization
b) The MD simulation is performed assuming a bound CO

Carbon ions are only formed through the complete atomization of the molecule. Assuming a uniform charge distribution at the ground state structure of the formic acid dimer, the MD simulations predict average KER values of 18.0 eV and 75.9 eV for $C^+$ and $C^{2+}$, respectively. These simulated values are in close agreement with the 17.7 eV and 85.8 eV measured in our experiment. However, the simulated KER predictions for oxygen exceed the 13 eV and 47.8 eV measured for $O^+$ and $O^{2+}$, indicating that a different mechanism is responsible for their production. By simulating the CE using a bound C=O fragment, the neighboring O receives a lower KER (of 13 and 49 eV) and the $O^+$ and $O^{2+}$ KER measurements agree with the MD simulations. Further, the simulated KER for all three ions of CO are also in agreement with our measured values. This suggests that the C atom has a strong tendency to remain bound in a CO formation, enabling $CO^+$ and $O^+$ to arise together with near equivalent signal intensity in the mass spectrum. Results of the simulation at a variety of degrees of ionization are presented in SI Tables S2-S5.

In summary, we have studied the Coulomb explosion and kinetic energy release from the ultrafast ionization and fragmentation of formic acid with fs laser intensities of $3.7\times10^{15}$ W/cm$^2$ at 400 nm. The production of $CO^{3+}$ was directly observed in the mass spectra and theoretical calculations show the existence of shallow excited state potential energy curves for $CO^{3+}$, confirming the existence of the metastable state. The lower limit for the trication lifetime determined through this method was on the order of a few hundred nanoseconds. MD simulations were performed and produce kinetic energy release values that are in agreement with the experimental measurements following the Coulomb explosion of formic acid dimer. We show that there is a strong preference for the CO to remain bound following multiple ionization. The approach outlined here could be applied to investigate other multiply charged fragment ions starting from the structures naturally found in larger molecules.

## ASSOCIATED CONTENT

Full experimental details, potential wells for $CO^{n+}$ (n=0-2) (Fig. S1) and parameters (Table S1), ionization rate calculations, peak splitting measurements (Fig. S2) and results from the MD simulation (Tables S2-S5) are available in the Supplemental Information.

## AUTHOR INFORMATION

**Corresponding Author**
*Scott.Sayres@asu.edu.

**Notes**
The authors declare no competing financial interests.

## ACKNOWLEDGMENT

We gratefully acknowledge support from ASU Lightworks.

Table of Contents artwork:

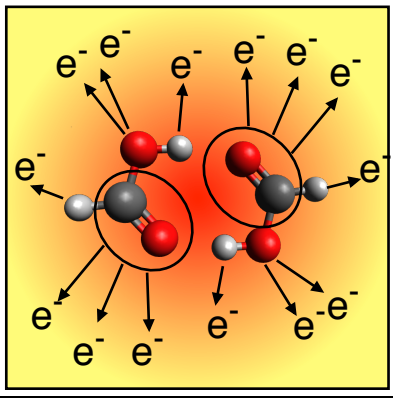

Supplemental Information for:

# Production of $CO^{3+}$ through the Strong-Field Ionization and Coulomb explosion of Formic Acid Dimer

Shaun F. Sutton,[1,2] Dane M. Miller,[1,2] Lenin M. Quiroz,[1,2] Ananya Sen,[1,2] Pilarisetty Tarakeshwar[1], and Scott G. Sayres[1,2]*

1 School of Molecular Sciences, Arizona State University, Tempe, AZ 85287

2 Biodesign Center for Applied Structural Discovery, Arizona State University, Tempe, AZ 85287

# 1. Experimental Methods

A home-built time-of-flight mass spectrometer housed in a vacuum chamber was used to explore the high energy ion fragments resulting from the Coulomb explosion of formic acid in the gas phase. Briefly, helium gas was bubbled through a liquid sample of (98% purity) formic acid and introduced to a vacuum chamber via pulsed nozzle. The gas traverses a skimmer to produce a molecular beam and travels 10 cm before reaching the extraction region of a Wiley-McLaren [1] time-of-flight mass spectrometer. Ionization of the neutral molecules occurs through interaction with a synchronized fs laser pulse. The ionized species are accelerated toward the detector with electrostatic grids that are held at static voltages (+4189 V and +2535 V). Cations travel through a 2 mm diameter skimmer, allowing for differential pumping to maintain a pressure of ~$1 \times 10^{-8}$ Torr in the detector chamber. Ions travel a 75 cm field-free region where they separate in time based on their mass/charge ratios before arriving at the MCPs. The signal is recorded by a digital oscilloscope and transmitted to a computer for analysis with an IEEE-488 interface. The low backing pressures (50 psi) used in this experiment only produces formic acid monomer and dimer as recorded in the mass spectra. Although the formation of larger clusters cannot be ruled out, they are not observed in the mass spectra and the KER measurements are consistent with the values expected from the dimer.

The Ti:sapphire laser system produces an output with wavelength centered at 800 nm, a power > 10 mJ per pulse, and operates at 30 Hz. The pulse width of the 800 nm laser beam was determined through an single-shot autocorrelator to be <150 fs. The 800 beam passes through a BBO crystal for second harmonic generation. 2 mJ of linearly polarized 400 nm light is directed into the chamber using a 50 cm focal lens, to reach laser intensities as high as $3.7 \times 10^{15}$ W/cm$^2$ and drive ionization and fragmentation of the molecular beam. The cross correlation of a non-resonant absorption ionization in Ar determined the pulse width of the 400 nm beam to be 200 fs.

## 2. Potential Energy curves

The potential energy curves for the multications of CO were calculated using CCSD(T) with an aug-cc-pVQZ basis set and are shown in Figure S4. The potential energy curves were fit to a Morse Well and fitting coefficients are presented in Table S1. The multication is considered metastable if the atomic energies are lower in energy than the bottom of the potential well and may live long enough to be observed with mass spectrometry. Ionization occurs via vertical transitions within the Franck-Condon region. We use the equilibrium bond length of CO to approximate the vertical IPs for comparison with literature assignments. The depth of the potential well of CO is calculated to be 11.1 eV, matching the literature value of 11.3 eV [2]. Our calculation shows the vertical IP of CO is 14.0 eV, in excellent agreement with literature values of 13.78 for the $^2\Sigma^+$ state [2].

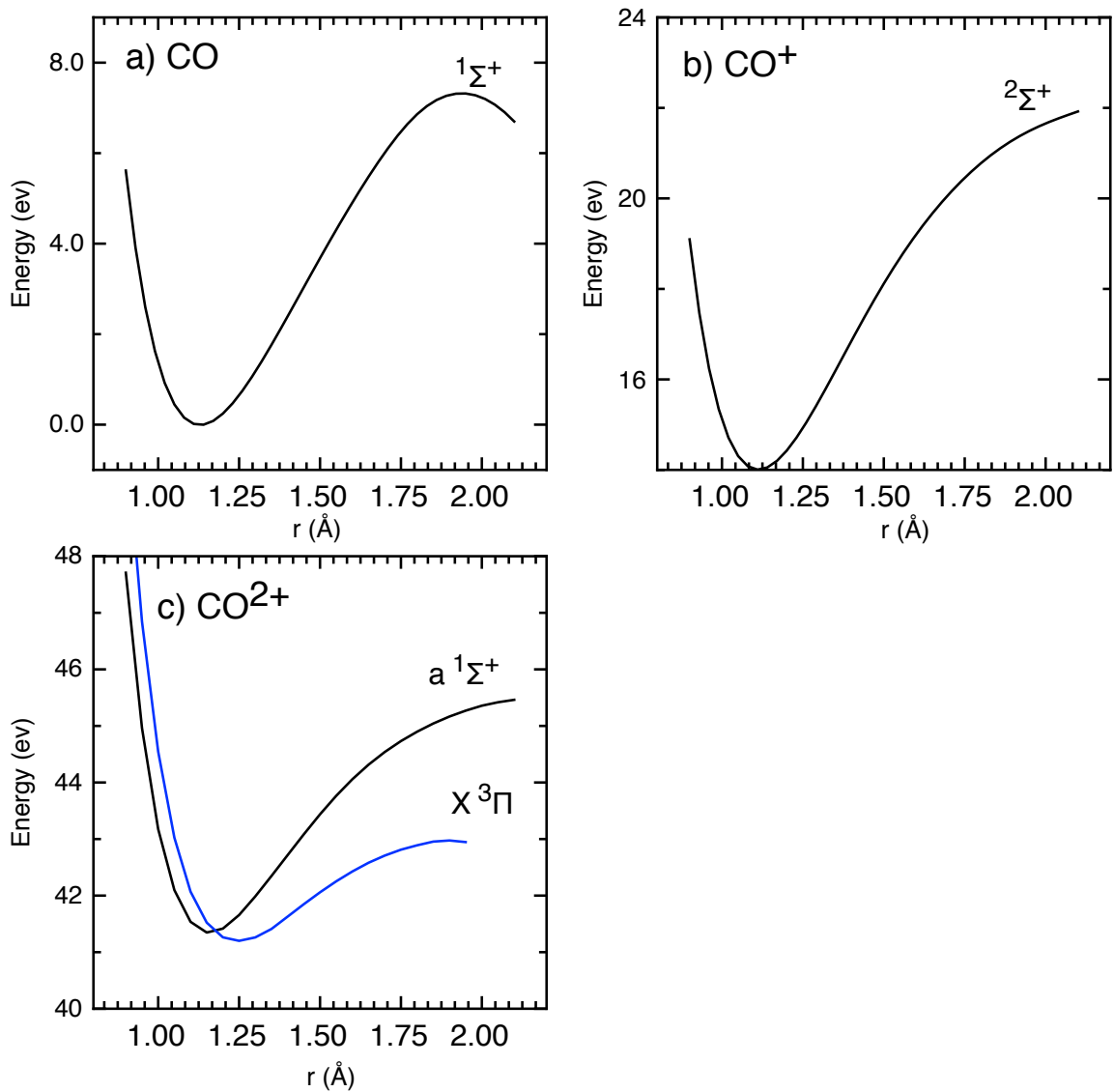

Figure S3. Calculated potential energy curves for the a) CO, b) CO$^+$, c) CO$^{2+}$ states of carbon monoxide using CCSD(T) with aug-cc-pVQZ.

Table S2. Potential well defining constants from CCSD(T) calculations.

| Ion Species | $D_e$ (eV) | $R_e$ | $\omega_e$ (cm$^{-1}$) | $\omega_e\chi_e$ (cm$^{-1}$) | $E_{vert}$ (eV) |
|---|---|---|---|---|---|
| CO $^1\Sigma^+$ | 11.1 | 1.130 | 2337 | 15 | 0.00 |
| CO$^+$ $^2\Sigma^+$ | 7.48 | 1.106 | 2499 | 20 | 14.03 |
| CO$^{2+}$ $^1\Sigma^+$ | 4.70 | 1.156 | 1979 | 25 | 41.4 |
| CO$^{2+}$ $^3\Pi$ | 1.77 | 1.249 | 1506 | 23 | 41.7 |
| CO$^{3+}$ $^2\Pi$ | 4.50 | 1.122 | 2299 | 20 | 83.6 |
| CO$^{3+}$ $^4\Sigma^+$ | 0.49 | 1.355 | 1137 | 23 | 82.23 |

## 3. Ionization Probability

The well-known ADK tunneling model [3] was used to calculate the ionization rate and is expressed in Eqs. (S1-S4). The stationary ionization rate ($W_{st}$) depends on the ionization potential ($E_i$), the laser's electric field amplitude ($E_0$), and the effective angular (l), magnetic (m), and effective principal (n*) quantum numbers, the charge state after the electron departs (Z), $\omega_0$ is the atomic frequency, $E_h$ is the atomic unit of energy, E is to convert the field of the laser to atomic units. The cycle averaged ionization rate (W) is computed from the ionization rate as shown in Eq. (S4).

$$W_{st} = \frac{\omega_o}{2} C_{n*l} \frac{E_i}{E_h} \frac{(2l+1)(l+|m|)!}{2^{|m|}(|m|)!(l-|m|)!} \left[2\left(\frac{E_i}{E_h}\right)^{3/2}\frac{E_0}{E}\right]^{2n*-|m|-1} \exp\left[-2/3\left(\frac{E_i}{E_h}\right)^{3/2}\frac{E_o}{E}\right] \quad (S1)$$

$$n* = Z/(E_i/E_h)^{1/2} \quad (S2)$$

$$C_{n*l} = \left(\frac{2e}{n*}\right)^{2n*}\frac{1}{2n*} \quad (S3)$$

$$W = \left[\frac{3}{\pi}\frac{E}{E_0}\left(\frac{E_h}{E_i}\right)^{3/2}\right]^{1/2} W_{st} \quad (S4)$$

With the assumption of a square wave laser pulse, the ionization probability, $P = 1-e^{-W\tau}$, can be used to show the laser intensities necessary to ionize the atoms composing the molecule, where $\tau$ is the laser's pulse width. The appearance intensity is determined by extrapolating the maximum slope of the curve back to the x-intercept of the laser intensity. This extracted number matches to an ionization probability of ~13%.

## 4. KER Measurements: Peak Splitting Method

A linear relationship between the ion momentum and the arrival time is produced when a Wiley McLaren [1] configuration is used. A typical signature of a Coulomb explosion is the production of two peaks for a given mass/charge. These features are produced as the ions are sent toward and away from the detector, relative to the flight direction within the applied static electric field. Larger KER distributions appear with an increased peak splitting. Experimental KER measurements for each ion are determined using the peak splitting method for time-of-flight mass spectrometry [4]. The kinetic energy release (KER) for each species was calculated as:

$$KER\ (eV) = \frac{0.1204 q^2 \Delta t^2}{m}\left(\frac{\Delta U_T}{d}\right)^2 \tag{S1}$$

where q is the charge of the ion, $\Delta t$ is the separation in time between the detection of the forward and backward ejected peaks in the mass spectrum, m is the molar mass, $\Delta U_T$ is the static field potential difference between the TOF grids, d is the spacing between grids, and the constant accounts for units. This measurement is commonly applied to atomic species, here, $CO^{2+}$ also demonstrates peak splitting as a result of KER because of the forces of the other ions composing the molecule. Peak splitting measurements for all high charge states are presented in Table 1 (and highlighted in Supplemental Figure S2).

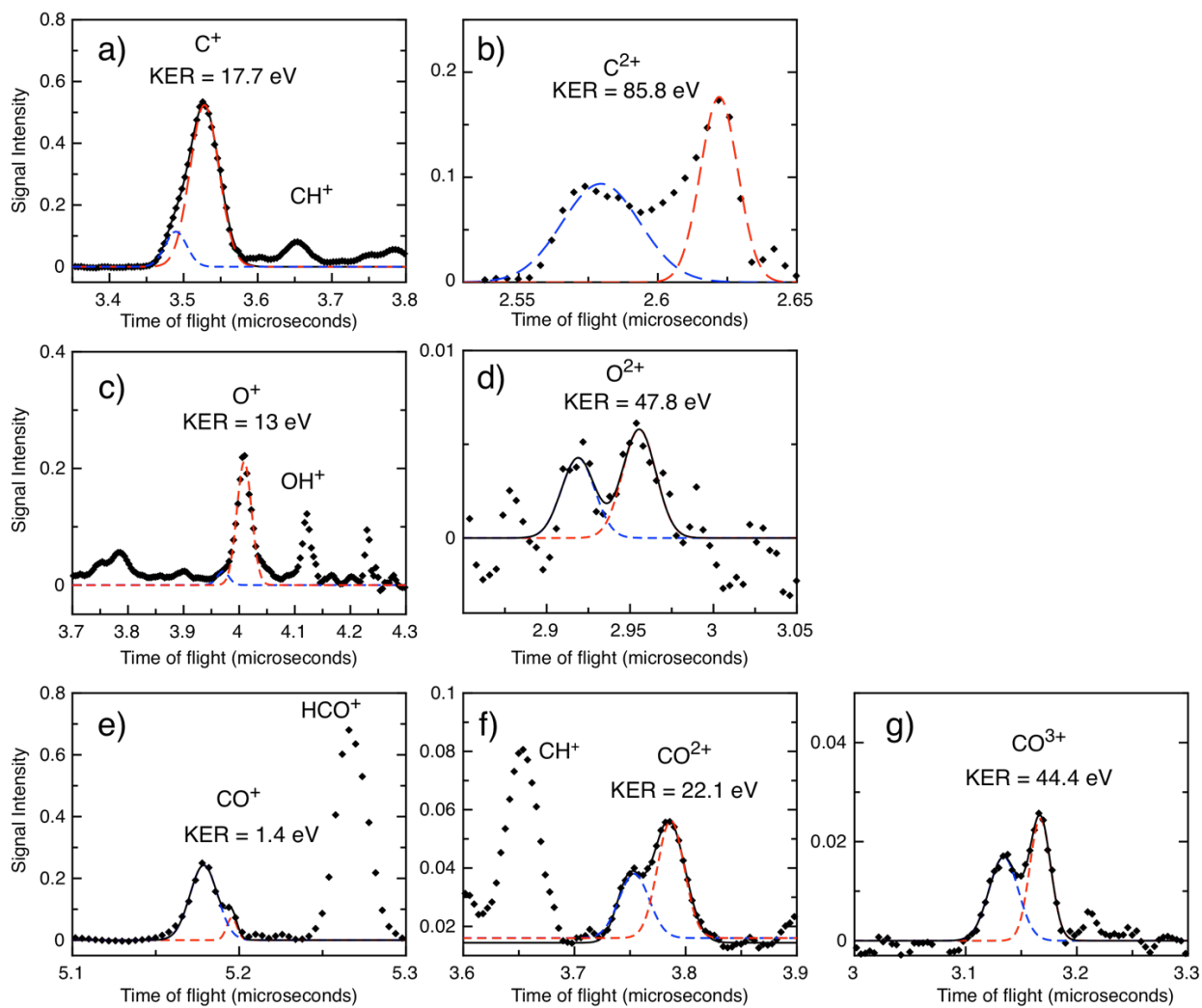

Figure S2. The peak splitting method is applied to each ion to measure the KER a) $C^+$ b) $C^{2+}$ c) $O^+$ d) $O^{2+}$ e) $CO^+$ f) $CO^{2+}$ g) $CO^{3+}$

## 5. Molecular Dynamics Simulations

The KER values are simulated using classical mechanics in a molecular dynamics simulation. These values are higher when considering the formic acid dimer and move closer to agreement with experiments. The magnitude of KER has a quadratic dependence on the charge states. Typically, these MD simulations are performed assuming individual atoms. However, to approximate the formation of an intact CO unit, simulations were performed with a ghost atom having a m/z of 28 and a coordinate at the center of mass of the CO. The intact CO causes minor changes in the trajectories and KER of the other ions.

The ground state structures of formic acid dimer were calculated with B3LYP/6-311++G(3d2f, 2p2d) level of theory using Gaussian16. [5] Within the formic acid dimer, the C-O bond length is 1.2431 and C=O is 1.3380 Å. The distance of the CO molecule is 1.1271 Å. [6] The cartesian coordinates for the atoms of the ground state structures of formic acid monomer and dimer are used as input for a MD simulation. The MD calculations provide a model to approximate the magnitude of the KER resulting from the Coulomb Explosion. For these simulations, we set the charge distribution of the clusters uniformly, according to the atomic IPs. The forces between the ions are computed under Coulomb's law and are allowed to act upon the ions for a small timestep (100 attoseconds), thereby accelerating them apart. After each step, the forces acting on the ions are recalculated using the new geometry. This process is repeated until the ions are sufficiently far apart to no longer influence the KER. Results from the MD simulation are shown in Tables S1-S4.

Table S2: MD simulation results for Formic acid molecule. All energies are provided in eV

| Number of Atoms | Mass | Charge | M/Z | Average KER | Max KER |
|---|---|---|---|---|---|
| 1 | 12 | 1 | 12 | 6.04 | 6.04 |
| 2 | 1.00784 | 1 | 1.008 | 30.439 | 30.678 |
| 2 | 15.999 | 1 | 15.999 | 11.62 | 13.016 |
| 1 | 12 | 2 | 6 | 28.468 | 28.468 |
| 2 | 1.00784 | 1 | 1.008 | 58.669 | 58.801 |
| 2 | 15.999 | 2 | 7.999 | 43.782 | 48.302 |
| 1 | 12 | 3 | 4 | 68.104 | 68.104 |
| 2 | 1.00784 | 1 | 1.008 | 87.331 | 88.477 |
| 2 | 15.999 | 3 | 5.333 | 96.172 | 105.202 |

Table S3: MD simulation results for Formic acid molecule with CO bound. All energies in eV

| Number of Atoms | Mass | Charge | M/Z | Average KER | Max KER |
|---|---|---|---|---|---|
| 1 | 28 | 1 | 28 | 3.272 | 3.272 |
| 2 | 1.00784 | 1 | 1.008 | 21.366 | 24.374 |
| 1 | 15.999 | 1 | 15.999 | 5.023 | 5.023 |
| 1 | 28 | 2 | 14 | 7.429 | 7.429 |
| 2 | 1.00784 | 1 | 1.008 | 29.131 | 30.923 |
| 1 | 15.999 | 1 | 15.999 | 9.086 | 9.086 |
| 1 | 28 | 2 | 14 | 12.67 | 12.67 |
| 2 | 1.00784 | 1 | 1.008 | 40.072 | 45.825 |
| 1 | 15.999 | 2 | 7.999 | 20.004 | 20.004 |
| 1 | 28 | 3 | 9.333 | 20.139 | 20.139 |
| 2 | 1.00784 | 1 | 1.008 | 47.778 | 52.4 |
| 1 | 15.999 | 2 | 7.999 | 28.74 | 28.74 |
| 1 | 28 | 3 | 9.333 | 28.095 | 28.095 |
| 2 | 1.00784 | 1 | 1.008 | 58.617 | 67.285 |
| 1 | 15.999 | 3 | 5.333 | 45.13 | 45.13 |

Table S4: MD simulation results for Formic acid dimer. All energies are provided in eV

| Number of Atoms | Mass | Charge | M/Z | Average KER | Max KER |
|---|---|---|---|---|---|
| 2 | 12 | 1 | 12 | 17.971 | 17.971 |
| 4 | 1.00784 | 1 | 1.008 | 48.586 | 52.167 |
| 4 | 15.999 | 1 | 15.999 | 15.342 | 17.866 |
| 2 | 12 | 2 | 6 | 75.664 | 75.664 |
| 4 | 1.00784 | 1 | 1.008 | 91.751 | 98.969 |
| 4 | 15.999 | 2 | 7.999 | 58.876 | 68.289 |
| 2 | 12 | 3 | 4 | 173.453 | 173.453 |
| 4 | 1.00784 | 1 | 1.008 | 136.266 | 146.649 |
| 4 | 15.999 | 3 | 5.333 | 130.448 | 150.907 |

Table S5: MD simulation results for Formic acid dimer with CO bound. All energies are in eV

| Number of Atoms | Mass | Charge | M/Z | Average KER | Max KER |
|---|---|---|---|---|---|
| 2 | 28 | 1 | 28 | 4.767 | 4.767 |
| 4 | 1.00784 | 1 | 1.008 | 34.115 | 38.527 |
| 2 | 15.999 | 1 | 15.999 | 12.587 | 12.587 |
| 2 | 28 | 2 | 14 | 17.5 | 17.5 |
| 4 | 1.00784 | 1 | 1.008 | 47.328 | 52.424 |
| 2 | 15.999 | 1 | 15.999 | 18.937 | 18.937 |
| 2 | 28 | 2 | 14 | 19.75 | 19.75 |
| 4 | 1.00784 | 1 | 1.008 | 61.64 | 70.461 |
| 2 | 15.999 | 2 | 7.999 | 49.048 | 49.048 |
| 2 | 28 | 3 | 9.333 | 41.858 | 41.858 |
| 4 | 1.00784 | 1 | 1.008 | 74.981 | 84.468 |
| 2 | 15.999 | 2 | 7.999 | 62.947 | 62.947 |
| 2 | 28 | 3 | 9.333 | 45.326 | 45.327 |
| 4 | 1.00784 | 1 | 1.008 | 89.022 | 102.177 |
| 2 | 15.999 | 3 | 5.333 | 109.172 | 109.172 |

## Supplemental References